# Modulating heat conduction by stretching or compressing

Jianjun Jiang and Hong Zhao*

Department of Physics and Collaborative Innovation Center of Chemistry for Energy Materials,

Xiamen University, Xiamen 361005, Fujian, China

*zhaoh@xmu.edu.cn

Recent studies have revealed that the symmetry of interparticle potential plays an important role in one-dimensional heat conduction problem. Here we demonstrate that by stretching or compressing the Fermi-Pasta-Ulam-$\beta$ lattice, one can control the symmetry of the potential, and thus manipulate the decaying behavior of the heat current autocorrelation function (HCAF). In fact, stretching or compressing induces a fast decaying stage (FDS) during which the HCAF decays faster than power-law manners or in a power law manner but faster than $\sim t^{-1}$. The time range as well as the decay amplitude of the HCAF over the FDS increase as the stretching or compressing ratio increase, or as the temperature decreases. As a consequence, the thermal conductivity calculated following the Green-Kubo formula shows a truncation-time independent window, implying a system-size independent conductivity. Stretching or compressing also changes the exponent of the power-law tail of the HCAF. The complicated heat conduction behavior induced by stretching or compressing can be connected to the change of the symmetry of the interparticle potential.

PACS number(s): *05.60.Cd, 44.10.+i, 05.20.Jj, 05.40.−a*

## I. Introduction

Heat conduction properties of one-dimensional lattices have attracted intensive studies for decades [1-34]. For one-dimensional momentum conserving lattices, a generally accepted viewpoint is that the heat current autocorrelation function (HCAF), defined as $C(t)=\langle J(t)J(0)\rangle$, decays in a power-law manner: $C(t)\sim t^{-\alpha}$, where $J(t)$ is the heat current fluctuations along the lattice at equilibrium [3-26]. As to the decaying exponent $\alpha$, though several values have been suggested following various theoretical analysis as well as numerical simulations, recent nonlinear hydrodynamic approaches show that it depends on the symmetry of the interparticle potential, i.e., $\alpha=1/2$ and $\alpha=2/3$, for symmetric and asymmetric potentials, respectively [19-21]. Numerical simulations support these predictions [22]. Following the Green-Kubo formula [35]

$$\kappa = \frac{1}{k_B T^2} \lim_{\tau_{tr}\to\infty} \lim_{L\to\infty} \frac{1}{L} \int_0^{\tau_{tr}} C(t)dt, \qquad (1)$$

the power-law decay implies the divergent thermal conductivity. Here $\kappa$, $T$ and $L$ are, respectively, the thermal conductivity, the temperature, and the size of the lattice. $k_B$ is the Boltzmann constant, which is set to be unity throughout. In Eq. (1), $\tau_{tr}$ is the truncation time of the integral. As the transport time of energy from one end of the lattice to another is proportional to the sound speed $v$, by truncating the Green-Kubo integral at $\tau_{tr}\sim L/v$ one can connect $\kappa$ to the finite-size nonequilibrium simulation results [13]. This treatment give rise to a size-dependent relation $\kappa\sim L^{1-\alpha}$, which has been confirmed by direct nonequilibrium simulations [3-10].

In spite of the prevailing viewpoint that heat conduction in one-dimensional momentum-conserving systems is anomalous, i.e., the heat conductivity diverges in the thermodynamical limit, some counterexamples that are momentum conserving but have convergent thermal conductivity, have also been reported [23-31]. To explain the inconsistence, one viewpoint is that the role the transient behavior of the HCAF plays in the thermal conductivity has been underestimated [27-30]. The power-law-decaying tail may appear in the thermodynamic limit, but before it shows up, the HCAF may first undergo a fast decaying stage (FDS), i.e., faster than power-law manners or in a power law manner but with $\alpha>1$, in a lattice with asymmetric potential. Under certain conditions, the FDS may last for a long time so that the HCAF decays over several orders. As a result, the contribution of the FDS to the Green-Kubo integral becomes dominant and the contribution of the tail is negligible. This mechanism can make the thermal conductivity size-independent practically, i.e., the observed conductivity appears as a constant from a short lattice with thousand particles to a lattice up to a macroscopic scale. In this sense, though the FDS is a finite-size effect as argued by some authors [32,33], it has practical importance for real applications.

Recently it is found that applying external pressure to the Fermi-Pasta-Ulam-β (FPU-β) lattice of symmetric interparticle potential can change its hydrodynamic transport behavior [21]. The scaling properties of the hydrodynamic modes with pressure are identical with that of the Fermi-Pasta-Ulam-αβ (FPU-αβ) lattice of asymmetric potential. Though anomalous diffusive behavior still remains, the effect of the pressure need be carefully investigated. In another recent work, by nonequilibrium simulations the authors have found hints of the convergent thermal conductivity in the FPU-β lattice when the external pressure is applied [34].

Indeed, studying the heat conduction behavior of a system under pressure has important practical significance, because materials subjected to such an environment do have certain interesting properties such as superconducting [36,37]. In particular, in recent years the effects of externally applied mechanical strain on heat conduction properties of the graphene has attracted much attentions due to the potential application importance [38-41]. However, the results reported in these works are very diverse and confusion; e.g., while some conclude that stretching or compressing can remarkably suppress the heat conduction [38,39], others state that the stretching may induce the divergence of the thermal conductivity [40,41]. The underlying mechanism has not been clearly understood.

In this paper, we show that stretching or compressing a one-dimensional lattice can change the asymmetry degree of its interparticle potential. Particularly, we show that the FPU-β lattice can be changed into a lattice of asymmetric potential by stretching or compressing. In more detail, we show that the HCAF in the FPU-β lattice decays in the power-law manner as predicted, but undergoes a FDS before the tail if being stretched or compressed. The FDS lasts longer if the stretching or compressing amplitude is increased or if the temperature is decreased. For sufficiently large stretching or compression and relatively low temperature, the HCAF may decay for several orders before the tail, making the thermal conductivity size-independent.

## II. Model and results

The Hamiltonian of the FPU-$\beta$ model is given by

$$H = \sum_i [\frac{p_i^2}{2} + U(x_i - x_{i-1} - 1)], \qquad (2)$$

with $U(x-1) = \frac{1}{2}(x-1)^2 + \frac{1}{4}(x-1)^4$, where $p_i$ and $x_i$ are respectively the momentum and position of the $i$th particle. Let $b \equiv (L-N)/N$ define the stretching or compression ratio, where $L$ is the lattice length and $N$ is the particle number. With the periodic boundary condition $x_{i+L}=x_i+L$, one can stretch ($b>0$) or compress ($b<0$) the lattice by adjusting $b$. Notice that $1+b$ represents the average inter-particle spacing and $x=1+b$ is the stationary state at zero temperature, inserting $x=1+b+\delta x$ into the potential $U(x-1)$ and expanding it around $x=1+b$ we get

$$U(b+\delta x) = U(b) + b(1+b^2)\delta x + (1+3b^2)(\delta x)^2 + b(\delta x)^3 + \frac{1}{4}(\delta x)^4. \qquad (3)$$

The first term in the right-hand side is a shift to the potential and therefore is trivial for the system dynamics. The coefficient $b(1+b^2)$ of the second term characterizes the pressure at zero temperature, and it is offset by the boundary restriction of fixed $L$. The last three ones are equivalent to the potential of the FPU-αβ lattice. At $b=0$, it returns to the symmetric FPU-β lattice with only symmetric terms of quadratic and quartic. In usual case, the cubic term leads to the potential asymmetric, and the ratio of this term to the quartic term can characterize the degree of asymmetry. Another important feature revealed from the expansion (3) is that the coefficients of the quadratic and quartic terms keep unchanged for $b=\pm\Delta$. The first term and the cubic term change only their sign under these transformations. This feature suggests that with the same ratio, the effects induced by the stretching and compressing may be identical. Our numerical simulations later will verify that the HCAF do show such property.

We evolve the lattice with the periodic boundary conditions for a sufficiently long time (t>10$^6$) to establish the equilibrium, and then we calculate the HCAF with 5x10$^8$ ensemble samples. The heat current is defined as $J(t) = \sum_i J_i$, with $J_i = \dot{x}_i \frac{\partial U}{\partial x_i}$. We fix the lattice number in our studies at N=8196. It is shown that the finite-size effect appears beyond $t \sim L/2v$ [20,22]. This time threshold represents the first encounter of the two sound modes at the boundary. It can be estimated that the sound velocity is about $v \sim 1+b$ at $T \sim 1$ in this lattice. With $N=8192$, we get a time window $t \in (0, 4098)$ in which the HCAF is not influenced by the finite-size effect. The second encounter of the sound modes takes place at $t \sim L/v$. Taking into account that the influence induced by a single collision of sound modes is small particularly when they encounter after a long time period, the HCAF till to the time length $t \sim 10^4$ is indeed can be considered as a good approximation to the infinite-size HCAF.

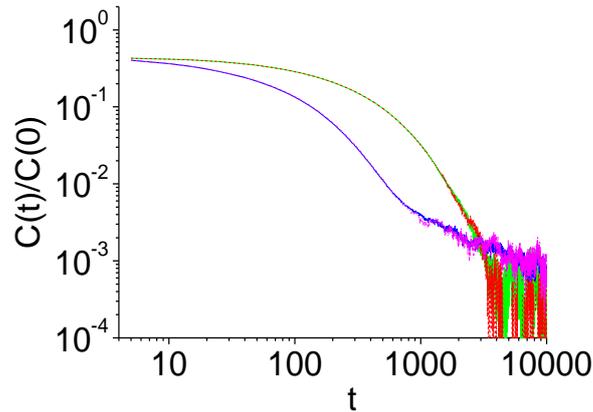

Fig. 1 Two pairs of HCAFs of b=-0.8 and 0.8 (two top curves), and b=-0.5 and 0.5 (two bottom curves) at T=1.

We first stress that the HCAF shows the identical decay behavior under the same ratio of stretching and compressing. Figure 1 show the log-log plot of HCAFs as functions of time at *b=-0.8, -0.5, 0.5* and *0.8,* respectively. The temperature is set at *T=1*. Obviously, pairs of HCAFs with *b=-0.8* and *0.8,* and *-0.5* and *0.5* overlap with each other, confirming that the HCAFs with b=±Δ are identical. Our studies below thus focus only on the case of stretching.

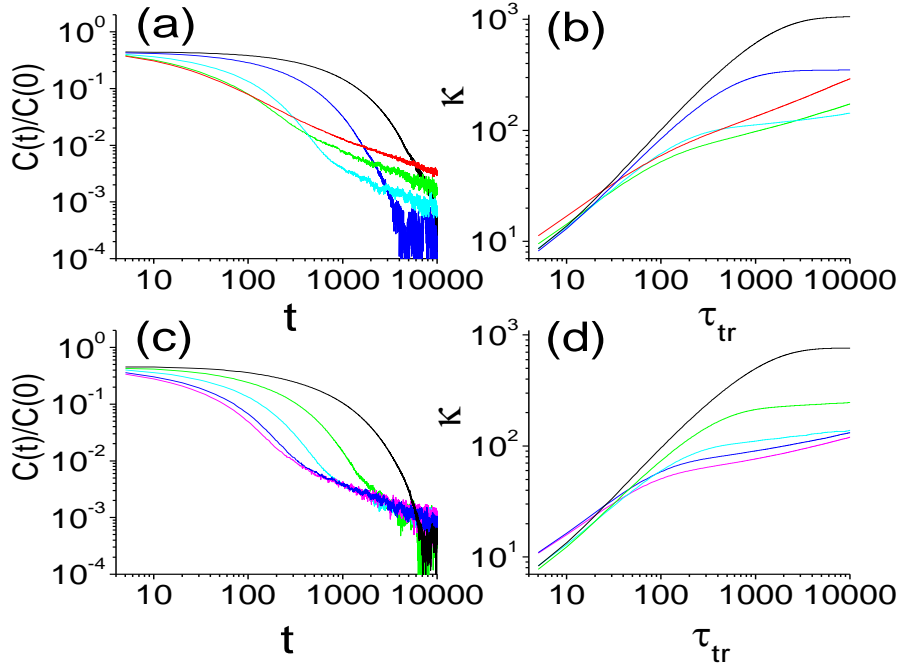

Fig. 2 (a) The HCAFs for b=1, 0.8, 0, 0.2 and 0.5 (from the above to the bottom at the line of t=1000) at T=1. (b) The thermal conduct calculated using HCAFs of Fig. 2(a) for b=1, 0.8, 0, 0.2 and 0.5 (from the above to the bottom at the line of t=10000) correspondingly. (c) The HCAFs for T=0.25, 0.5, 1, 2, and 3 (from the above to the bottom) at b=0.5. (d) The thermal conductivity calculated using HCAFs of Fig. 2(c) for T=0.5, 1, 2 and 3 (from the above to the bottom at the line of t=1000) correspondingly.

We then report main results. Figure 2(a) shows the log-log plot of the HCAF as a function of time at $b=0, 0.2, 0.5, 0.8,$ and $1$ with temperature $T=1$. At $b=0$, the HCAF decay in a perfect power-law manner in the tail part. In other cases, each HCAF shows up a FDS in the initial stage, and the scale of the FDS increases with the increase of $b$. Following the FDS, a power-law tail appears definitely in the cases of $b=0.2$ and $0.5$. At $b=0.8$ and $1$, however, the statistical accuracy of the data does not allow for a conclusive judgement to whether the tail exists. The HCAFs have decayed over three orders with such two stretching ratios.

To probe the possible tail in such low amplitudes one needs much more ensemble samples, which beyond our computers ability since roughly 100 times of more ensemble samples can suppress the fluctuation amplitude of the HCAF by only one order. This is nevertheless not a serious problem for estimating the thermal conductivity, since after several orders decay the contribution of the possible tail for $\kappa$ could be negligible small. To explain this argument, we divide the Green-Kubo integral into two parts, $\kappa = \frac{1}{k_B T^2 L}[\int_0^\tau C(t)dt + \int_\tau^{\tau_{tr}} C(t)dt] = \kappa(\text{FDS}) + \kappa(\text{tail})$, by the turning point $\tau$ from the FDS to the power-law decaying stage. In the case of $b=0.8$, for example, assuming that the tail of $C(t) \sim t^{-2/3}$ appears following the FDS, it has $\kappa(tail)/\kappa(FDS) \sim 0.5$ if truncating the integral at $\tau_{tr} \sim 10^8$. In more detail, the observed $\kappa$ keeps as a constant approximately from $\tau \sim 10^4$ to $\tau_{tr} \sim 10^8$. The lattice should reach up to $N \sim vt \sim 10^8$ to evidence a *50%* increase. This system size already reaches macroscopic sizes of real 1D materials with about $10^8$ molecules on a line.

Figure. 2(b) shows κ calculated by the Green-Kubo formula as a function of the truncation time $\tau_{tr}$ using HCAFs of Fig. 2(a) correspondingly. Since $\tau_{tr} \sim L$, it also represents the system-size dependent behavior of $\kappa$. The behavior of $\kappa$ is determined by the scale of the corresponding FDS. There is no FDS in the FPU-*β* lattice. The HCAF decays in a power-law manner $C(t) \sim t^{-\alpha}$, though in the initial stage there is a region that it decays faster than in the tail part but the exponent remains as α<1. As a result, κ diverges in a power law manner. At *b=0.2*, the FDS appears in a short period. Fitting the end part of the FDS with $C(t) \sim t^{-\alpha}$, it gives α~1.2. Therefore, the influence of the FDS to κ is not significant, though it already decreases the increase rate comparing to the case of the FPU-*β* lattice. At *b=0.5*, a relatively large FDS appears which leads to κ converge to a plateau at large times. In this case, the FDS is still not large enough. The tail starts to contribute the integral eventually and induces a slight increase of κ in the plateau window. In the cases of *b=0.8* and *1*, the FDSs become sufficient large and κ converge to time-independent constants. One can expect its re-increase but after $\tau_{tr} \sim 10^8$ as argued in above paragraph.

Comparing to the FPU-*β* lattice, stretching with *b<0.5* suppresses the heat conduction. While with extreme stretching, κ converges to the time-independent plateau with very high value. If measuring the thermal conductivity of short lattices, stretching enhances the heat conduction. But because κ increases in the FPU-*β* lattice in the power law manner, finally it will exceed the plateau value. Therefore, for large enough lattices, the conclusion is that the stretching always suppresses the heat conduction of the FPU-*β* lattice.

If making comparison among the stretched lattices, the situation appears more complicatedly. Under the extreme stretching, as the cases of *b=0.8* and *b=1*, one may conclude that stretching enhances the heat conduction. With relatively small stretching, this conclusion keeps correct for shorter lattices. While because κ increases faster with smaller stretching ratio than with bigger ones, the conclusion may be reversed for large enough lattices, i.e., stretching suppresses the heat conduction. However, when the conclusion is reversed depends on the lattice length, the stretching ratio.

Figure 2(c) shows HCAFs at several temperatures, *T= 0.25, 0.5, 1, 2* and *3*, respectively, at *b=0.5*. The FDS increases with the decrease of temperature. Tails following FDSs seem to collapse together with each other. The corresponding κ calculated by the Green-Kubo formula with the integral truncated at $\tau_{tr}$ are shown in Figure 2(d), revealing that even with relatively small compression ratio the time-independent plateau on κ may still present at lower temperatures. The dependence of the κ on the temperature is also complicated, which depends on the lattice size that κ is measured.

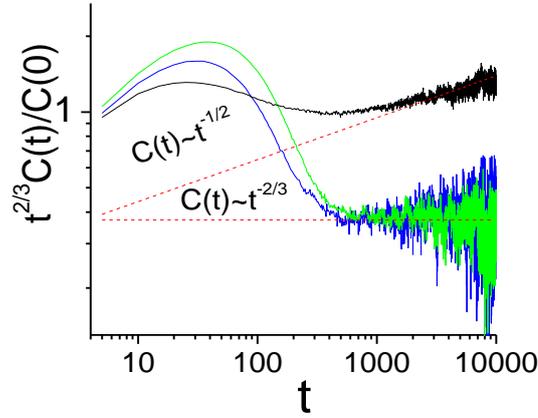

Fig. 3 The rescaled HCAFs for *b=0* at *T=2* (the top curve), and *b=0.5* at *T=2* (the mid curve) and *T=3* (the bottom curve),

We finally discuss the tail following the FDS. In the case of big stretching ratio and lower temperatures, we could not concrete the exact decay manner of the tail because of the statistical precision of data. In other cases, definite power-law tails can be identified. To check the decay exponent we plot $t^{2/3}C(t)$ as a function of time in Fig. 3 at *b=0.5* with *T= 2* and *T=3* respectively. It is clear that the tails match the theoretical prediction of $C(t) \sim t^{-2/3}$ well for asymmetric interparticle potentials. For the purpose of comparison, the HCAF of the FPU-β lattice at $T=2$ is also plotted in the figure with the same rescaling factor $t^{2/3}$. It shows that $C(t)$ does not fit the $t^{-2/3}$ law. The best fitting to the tail gives $\alpha \sim 0.55$ indeed. This value diverges slightly from the theoretical prediction of *1/2*, which should be due to the insufficiency of the lattice size. It has shown that α in the case of the FPU-β lattice will converge to the predicted value until up to about $N \sim 10^5$ [22]. Therefore, we may conclude that the stretching/compression converts the decay exponent of the HCAF tail from *α=1/2* to *α=2/3*.

### III. Summary and discussions

In summary, stretching or compressing the FPU-β lattice can change its heat conduction behavior dramatically. They change the symmetric interparticle potential to be asymmetric and result in an initial FDS in the HCAF. The time range of the FDS and the decay of the HCAF over the FDS depend on the stretching or the compressing ratio, the temperature and the size of the system. Following the FDS, a power-law tail may appear, particularly when the stretching or the compressing ratio is relatively low and the temperature of the system is high. In this case the asymmetric potential changes the decaying exponent of the tail to be α=2/3. In contrast, if at a given temperature the stretching or the compressing ratio is big enough or at a given stretching or compressing ratio the temperature is low enough, the HCAF may decay over several orders within the FDS, making the contribution of the tail to the heat conductivity negligible even for a macroscopic system size. In this case, the thermal conductivity is in effect system-size independent.

To analyze the potential symmetry change and thus to precisely predict how it affects the system's heat conduction behavior, it is important to expand it at the stationary state of the system under stretching or compressing. For the FPU-β lattice we have studied, the expanded potential consists of a quadratic, a cubic, and a quartic term, among which the cubic term is introduced by the stretching or compressing. The amplitudes of these terms are unchanged under the transformation from *b* to *–b*, and thus the effects induced by the stretching and compressing with the same ratio are identical. Our simulations have verified that the HCAFs of *b* and *–b* overlap with each other. The dependences of coefficients of these terms on the stretching or the compressing ratio are not synchronous. At a fixed temperature, increasing the stretching or the compressing ratio amplifies the coefficients of the quadratic and cubic terms, but the amplifying ratio is bigger for the former than the latter. Meanwhile, the coefficient of the quartic term keeps unchanged. As a consequence, the harmonic interaction becomes increasingly important, while the asymmetric, cubic term becomes increasingly dominant to the unharmonic interactions. The former effect tends to slow down the decay of the HCAF since it drives the system close to the integrable limit but the latter tends to change the decay manner of the HCAF. This explains why in the initial stage though the HCAF decays slowly but in an exponential-like manner. This mechanism for the formation of the FDS works equally when one changes the temperature at a fixed stretching or compressing ratio.

**Acknowledgments**

This work is supported by the National Natural Science Foundation of China (Grants No. 11335006).